# CHO-bearing molecules in Comet 67P/Churyumov-Gerasimenko

*Markus Schuhmann[1]\*; Kathrin Altwegg[1,2]; Hans Balsiger[1]; Jean-Jacques Berthelier[3]; Johan De Keyser[4]; Stephen A. Fuselier[5,6]; Sébastien Gasc[1]; Tamas I. Gombosi[7]; Nora Hänni[1]; Martin Rubin[1]; Thierry Sémon [1]; Chia-Yu Tzou[1]; Susanne F. Wampfler[2]*

(1) Space Research and Planetary Sciences, University of Bern

(2) Center of Space and Habitability, University of Bern

(3) LATMOS/IPSL, Sorbonne Université, Paris

(4) Royal Belgian Institute for Space Aeronomy, Brussels

(5) Department of Space Science, Southwest Research Institute

(6) University of Texas at San Antonio

(7) Department of Climate and Space Sciences and Engineering, University of Michigan

ABSTRACT

In 2004, the Rosetta spacecraft was sent to comet 67P/Churyumov-Gerasimenko for the first ever long-term investigation of a comet. After its arrival in 2014, the spacecraft spent more than two years in immediate proximity to the comet. During these two years, the ROSINA Double Focusing

Mass Spectrometer (DFMS) onboard Rosetta discovered a coma with an unexpectedly complex chemical composition that included many oxygenated molecules. Determining the exact cometary composition is an essential first step to understanding of the organic rich chemistry in star forming regions and protoplanetary disks that are ultimately conserved in cometary ices. In this study a joint approach of laboratory calibration and space data analysis was used to perform a detailed identification and quantification of CHO-compounds in the coma of 67P/Churyumov-Gerasimenko. The goal was to derive the CHO-compound abundances relative to water for masses up to 100 u. For this study, the May 2015 post-equinox period represent the best bulk abundances of comet 67P/Churyumov-Gerasimenko. A wide variety of CHO-compounds were discovered and their bulk abundances were derived. Finally, these results are compared to abundances of CHO-bearing molecules in other comets, obtained mostly from ground-based observations and modelling.

Keywords: Jupiter-Family Comets; Rosetta; Organics; Mass Spectrometry, Solar System; Space Missions;

## 1. Introduction

Until the Rosetta mission, space missions targeting comets were generally brief fly-bys through the coma or tail. As a consequence, although comets are considered to contain many important details about the evolution of our solar system, much of this information was yet to be revealed. Most data on the chemical composition of cometary nuclei and comae were obtained using remote sensing methods from ground. While ground-based measurement techniques have the advantage of being able to observe many cometary comae, these techniques are usually limited to chemical compounds with relatively large dipole moments. For compounds with small dipole moments, these techniques result in large uncertainties in the molecular abundances.

The Rosetta spacecraft, launched in 2004, followed a new approach of long-term encounter and close-up investigation of comet 67P/Churyumov-Gerasimenko (hereafter 67P). The spacecraft followed the comet throughout most of its orbit around the Sun, thus including observations during essential events like the in- and outbound equinoxes and perihelion. With Rosetta, new data about comets were collected, with many of the physical and chemical properties revealed for the first time.[1,2] For example, the complex shape of the nucleus, its very dark surface, and the huge variability in the coma were surprising. Further, the richness and complexity of the comet's organic composition were detailed during the two-year encounter. Although a fly-by of the Giotto spacecraft already showed the presence of organic molecules up to 100 u/e in comets [3], the increased mass resolution and sensitivity of the instruments onboard Rosetta revealed a diversity of complex organic molecules far beyond expectations. One year after arrival at the comet, Le Roy et al.[4] published a chemical inventory of molecules found early in the mission, at a solar distance of 3 au, confirming the presence of many organic compounds previously only detected via remote ground-based methods.

The instruments onboard the Rosetta spacecraft included several mass spectrometers. The lander Philae had two mass spectrometers, COSAC [5] and Ptolemy [6], which collected information directly from the comet surface. Onboard the Rosetta Orbiter, the mass spectrometer COSIMA studied the composition of dust grains in the coma, while ROSINA mainly observed the gaseous phase of the coma. The ROSINA suite included two mass spectrometers that observed the coma at a wide range of distances from the comet surface throughout the two-year encounter. All groups of mass spectrometers performed studies on the chemical composition, and their combined observations exhibited both similarities and differences. Altwegg et al.[7] provided a detailed summary of these studies, including discussion of the presence of CH-, HCN-, and CHO-compounds in the comet. In addition, a detailed study on hydrocarbons across several mission phases was performed by Schuhmann et al.[8], showing the presence of a range of aliphatic and aromatic compounds in the coma.

In this study, a calibration campaign was performed on several CHO-bearing molecules to obtain their fragmentation patterns and the detector sensitivity for ROSINA-DFMS mass spectrometer. This campaign forms the basis of a detailed identification and quantification of CHO-bearing molecules up to 100 u/e in the coma of comet 67P. Furthermore, by combining the laboratory and space data, the bulk abundances of CHO-bearing molecules relative to water are determined. Data presented here were compared to results from other cometary studies, as well as modelling results of the interstellar medium. Section 2 describes the instrumentation and the laboratory calibration. Section 3 applies this laboratory calibration to the space data. Section 4 discusses the conclusions.

## 2. Methods and Observations

The ROSINA instrument suite onboard of the Rosetta spacecraft consists of three different sensors. The COmetary Pressure Sensor (COPS), measuring the overall density, the Reflectron-type Time-Of-Flight Mass Spectrometer (RTOF), and the Double Focusing Mass Spectrometer (DFMS).[9] This work focuses on data obtained with DFMS. Consisting of a Mattauch-Herzog-configuration[10], DFMS has a high mass resolution (3000 at 1% peak height) and a high sensitivity, which together with the instrument's long operation time onboard of Rosetta provide ideal conditions for detailed analysis of the comet's organic composition. DFMS makes measurements in high- and low-resolution mode. However, for this study high-resolution mode data are used exclusively. There are two nearly exact copies of DFMS: the flight model onboard of the spacecraft and a laboratory model with nearly identical properties and identical settings to the flight model. All calibrations discussed below were performed with the laboratory model, while the flight model collected space data.

2.1 Laboratory Calibration

For calibration the University of Bern calibration facility CASYMIR (Calibration System for the Mass Spectrometer Instrument ROSINA) was used. This facility was developed for measurements under space-equivalent conditions. In particular, it is operated in very low density conditions from $10^{-10}$ to $10^{-6}$ mbar.[11,12] For calibration of DFMS, a series of measurements were performed at three different pressure levels, ranging from $10^{-9}$ to $10^{-7}$ mbar, to derive the fragmentation behavior of the compounds and the sensitivity of the instrument to the compounds.

2.1.1 Calibration of Fragmentation

DFMS uses electron impact ionization to ionize very low energy cometary molecules. Besides charging neutral molecules positively, molecular chains may be fragmented into smaller pieces of unsaturated or even saturated species through this electron bombardment. Simulation and pre-flight testing of the DFMS ion source showed that the best performance was achieved at an ionization energy of 45 eV. This energy is lower than the 70 eV ionization energy traditionally used in the laboratory to develop molecular fragmentation databases. Fragmentation patterns of molecules are affected by the ionization energy. The patterns directly obtained in the DFMS calibrations were thus the primary source for this study. More traditional databases provided fragmentation patterns as a secondary option when DFMS calibration measurements could not be performed. There were several reasons why some calibration measurements could not be performed: (I) Compounds had vapor pressures that were too low, so that vacuum conditions applied would not be sufficient for creating the level of gas phase required for calibration. (II) Compounds vapor pressures were too high, in some cases leading to unstable pressure conditions, difficult to correct with the thermal valve control elements used, or (III) Compounds were corrosive and/or poisonous and therefore too dangerous to study. For compounds not measured in the DFMS, the Database of the National Institute for Standards and Technology [13] (further referred to as NIST) were used.

DFMS fragmentation patterns are obtained either relative to the total amount of ions or to the most abundant fragment. An example of the fragmentation behaviour of methanol in DFMS is provided in Table 1 with the corresponding fragmentation pattern in Figure 1. A list of all fragmentation values used for this study is found in the appendix.

High quality fragmentation patterns are essential for identification of molecules in the DFMS space data, as fragmentation processes can also lead to formation of molecules present in the volatile phase of the comet (e.g. $CH_2O$ from $CH_3OH$). This substantially complicates the identification of CHO-bearing molecules in the coma. Only molecules showing at least the parent and major fragments, as determined from the lab calibration, are classified here as present in the space data. In addition, any impact from other compounds sharing fragments of the same mass must be subtracted according to the fragmentation pattern.

2.1.2 Calibration of DFMS Sensitivity

The DFMS detector consists of a microchannel plate (MCP) with a position sensitive linear anode (Linear Electron Detector Array, LEDA) that has a resolution of 512 pixels. Detailed information on the detector is found in Nevejans et al.[14] and Balsiger et al.[9]. The detector together with the analyzer show a species (mass)-dependent sensitivity for which a compensation factor, the so-called sensitivity factor (Appendix A), is derived from calibration measurements. This factor (given in $cm^3$) relates the density of the compound in DFMS's ion source multiplied by the ratio of the electron emission current to the ion current on the detector at each pressure level. A detailed description of the DFMS sensitivity derivation is found in Le Roy et al.[4]

**Table 1: DFMS fragmentation of methanol**

| Mass (u) | Fragment | Abundance (%) | Error (%) |
|---|---|---|---|
| 13 | CH | 4.05 | 1.2 |
| 14 | $CH_2$ | 12.60 | 3.6 |
| 15 | $CH_3$ | 100.00 | 0.0 |
| 16 | O | 0.52 | 0.2 |
| 17 | OH | 1.23 | 0.4 |
| 28 | CO | 14.89 | 4.3 |
| 29 | CHO | 50.09 | 9.8 |
| 30 | $CH_2O$ | 7.22 | 2.7 |
| 31 | $CH_3O$ | 63.12 | 11.9 |
| 32 | $CH_4O$ | 39.18 | 9.7 |
| 33 | $^{13}CH_4O$ | 0.52 | 0.2 |

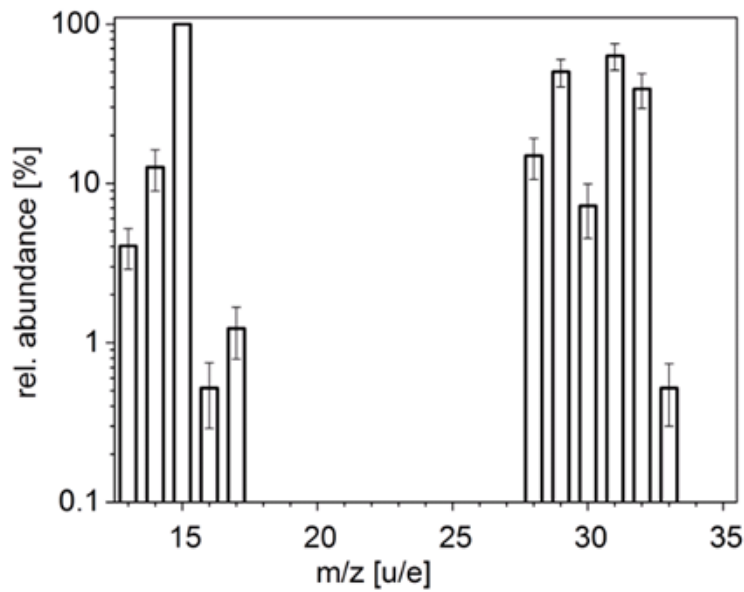

**Figure 1: DFMS fragmentation pattern of methanol.** Bars represent the degree of fragmentation relative to the most abundant fragment ($CH_3$).

All calibration compounds are present in liquid phase at atmospheric pressure and room temperature, thus requiring a transformation into gaseous phase for the DFMS measurements. In the calibration chamber, samples were exposed to vacuum ($< 10^{-4}$ mbar) at temperatures of < 353.15 K, allowing higher evaporation rates. For this study, calibration of a broad range of CHO-bearing molecules was performed. However, as the calibration capacity is limited, the main focus was on CHO-compounds that were detected in the coma of the comet [cf 7] and on compounds expected to be present in the coma due to their presence in the interstellar medium or in other comets.

Data obtained from separately performed background measurements were subtracted from the calibration output, removing any contribution and contamination from previous calibration runs in CASYMIR or in the instrument. Sensitivity is derived from all ions detected as a function of pressure level. Sensitivity calculation for three different pressure levels showed a linear correlation. This correlation is approximated using least-square fit in the form of a linear regression, with the slope representing the sensitivity factor.

2.2 Space Data Selection

For this work, DFMS space data from the May 2015 post-equinox period (date: May 21 to June 1, 2015; mission phase: ESC2–MTP016, STP058 - VSTP111) were selected. This mission phase was considered to be the best representation of the organic bulk composition of comet 67P. At a heliocentric distance of 1.53 au, the comet was considered close enough to the Sun to enable a sufficient outgassing rate, whilst unstable outburst conditions such as those occurring around perihelion could be avoided. The distance between the comet and the spacecraft was relatively large (~200 km), which is known to have an impact on the DFMS spectra intensities. For the derived abundances, the time period from May 21 to June 1 was used, as the spacecraft was located over the comet's southern (summer) hemisphere. Data obtained during spacecraft maneuvers were excluded as previous studies indicate that maneuvers lead to the desorption of larger amounts of molecules from the spacecraft surface.[15]

2.3 Data Treatment

The DFMS raw spectra consist of 512 abscissa points (representing the LEDA pixels) with the number of detector counts at each pixel. DFMS data analysis requires a use of a mass scale (m/z) and a calculation of the number of ions from the detector counts. The number of ions is generally calculated over 20s, referring to the default 20s integration time. Also, an individual pixel gain is applied, compensating for local variation in the sensitivity of the LEDA pixels. In addition, a general gain and a mass dependent detector yield value are applied. The number of ions per species is derived from integration of the peak over each pixel. However, in this study a single-Gaussian fitting routine was used to fit the peak shapes (fig.2) as this represents the core of the peak very well and is a simpler and preferable procedure to separate overlapping peaks in spectra when compared to summing the number of ions over each pixel around the peak.

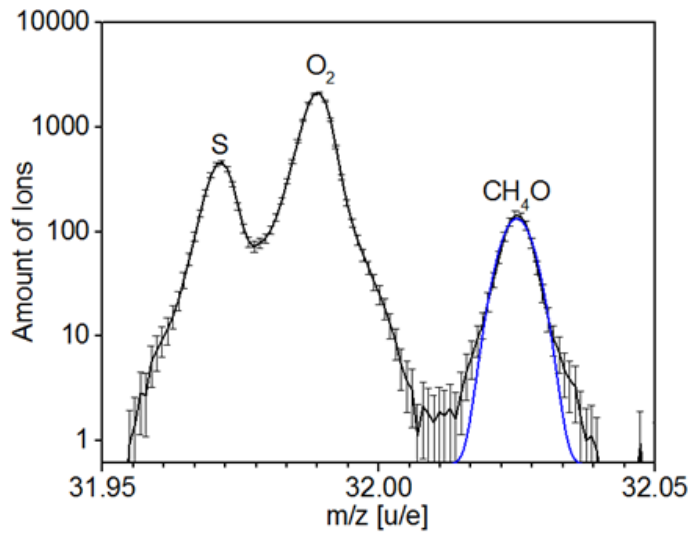

**Figure 2: CH$_4$O peak in DFMS high-resolution spectra.** For analysis of the CHO-bearing molecules a single-Gaussian fitting routine was applied in the spectra.

The relative abundances of molecules were calculated using both the DFMS laboratory and space data. Therefore, fragmentation and sensitivity, as obtained in calibration, and the number of ions per species, as obtained from the space data, are taken into account. The procedure for computing the abundance ratios is described in detail for DFMS by Gasc et al.[16]

## 3. Application to Space Data

3.1 Presence of CHO-species lighter than 75 u/e

Analysis of the DFMS space data from May 2015 inbound post-equinox reveals the presence of several CHO-bearing compounds. High amounts of $CH_2O$ and $CH_4O$ are found in the DFMS data at 30 u/e (fig.3.a) and 32 u/e (fig.3.b) and can be identified as methanol ($CH_3OH$) and formaldehyde ($H_2CO$). Formaldehyde is also a fragmentation product of methanol; however, as shown in Figure 1, the level of fragmentation of methanol into formaldehyde is not high enough to explain the large amounts present. At 44 u/e, smaller amounts of $C_2H_4O$ can be found (fig.3.c) and identified as acetaldehyde ($CH_3CHO$), while $CH_2O_2$ and $C_2H_6O$ are present at 46 u/e (fig.3.d), representing formic acid (HCOOH) and ethanol ($C_2H_5OH$). In terms of fragmentation, acetaldehyde and ethanol show a pattern similar to methanol and formaldehyde, that is, a loss of two single-bonded H-atoms under electron impact causes formation of $C_2H_4O$ from $C_2H_6O$. But again, this process is not sufficient to explain the presence of all $C_2H_4O$ and the high levels of $C_2H_4O$ essentially confirm that acetaldehyde is present in the coma in May 2015. The peak at 42 u/e is identified as $C_2H_2O$. However, in this case, a fragmentation contribution from almost all CHO-bearing molecules of higher mass present is possible. Therefore $C_2H_2O$ is categorized here as a fragmentation product. For the selected measurement period, the presence of propanol and butanol is considered to be likely, as data from other periods shows propanol and butanol in the coma of the comet 67P. Le Roy et al.[4] and Altwegg et al.[7] confirmed the presence of propanol. However, neither propanol nor butanol were detected in this study from only the May 2015 data. The failure to detect butanol here is considered to be due to low abundances of both molecules and the large distance between the comet and the spacecraft, resulting in lower intensities. Both species are thus expected to be present, but they are considered hidden in the background and are excluded from the calculation of the relative abundances. At mass 58 u/e, $C_3H_6O$ is present (fig.3.e). For this peak, identification based on structural features is difficult as several isomers could be present. The most feasible candidate is probably acetone, in agreement with Altwegg et al.[7], who identified acetone as well. Furthermore, the presence of acetone in the interstellar medium and on comets is very well documented, and therefore in contrast to other isomers like propanal that could be present. Differentiation of both compounds via fragmentation pattern should be possible. Propanal produces high quantities of fragmentation products around 29 u/e, while acetone shows only very little fragmentation around this mass/charge ratio. Instead, for acetone, high quantities of fragmentation products occur at 43 u/e. Comparing DFMS spectra at 29 u/e and 43 u/e from May 2015, the CHO-peak at 29 u/e appears to be higher than the peak at 43 u/e by roughly a factor 10. Hence, the fragmentation values and the sensitivity of propanal were used for calculation of the relative abundance. However, the peak at 29u/e has contributions from fragmentation from all major CHO-bearing compounds present, hence doesn´t allow unequivocal isomer identification. Most likely there is a mixture of both (or more) isomers and the exact ratio must remain undetermined.

A peak at 60 u/e refers to another CHO-bearing compound (fig.3.f). It confirms the presence of acetic acid, which proves together with above discussed formic acid, the existence of carboxylic acids in the coma of the comet 67P. At 62 u/e, low amounts of $C_2H_6O_2$ are present (fig.3.g), potentially belonging to ethylene glycol.

Another peak at 72 u/e (fig.3.h) is identified as $C_4H_8O$. For this peak, the fragmentation pattern and sensitivity of butanal were used in the analysis. Again, one of several isomers, or a combination of them, can be responsible for the peak. In addition to butanal, there is the possible presence of its isomers butanaone and propanal, 2-methyl- (isobutanal). The NIST fragmentation pattern of butanal and isobutanal are similar. Besides the rather small differences in the fragmental abundances, the only major difference is the lack of a fragmentation product at 44 u/e in the isobutanal-pattern. Hence, differentiation between these isomers is not possible under the present circumstances.

Another peak at 74 u/e (fig.3.i) is identified as $C_3H_6O_2$, which would be a known fragmentation product of propylene glycol. However, propylene glycol could not be identified in this study (see following section), instead there may be the presence of methyl acetate or an isomer such as methyl formate at 74 u/e. Methyl formate is a good candidate because of its presence in the ISM. However, the NIST fragmentation pattern of methyl formate shows a different picture, as no observable parent peak or peaks of close-by fragments are present in the fragmentation pattern. In contrast, the NIST fragmentation pattern of methyl acetate clearly indicates the presence of a parent peak. As this is in agreement with the DFMS space data, the fragmentation pattern and sensitivity of methyl acetate were used for this study.

3.2 Presence of CHO-Species Higher than 75 u/e

Species of higher masses are difficult to investigate as fluxes from the comet decrease with increasing masses and the instrument sensitivities decline. Besides that, the larger distances between the comet and the spacecraft in May 2015 result in rather lower intensities compared to other mission phases. Nevertheless, the identification campaign was performed up to 100 u/e. The conditions for detection of glycolic acid and propylene glycol were considered to be especially favorable in May 2015 compared to other periods when strong contributions of $C_2H_4OS$ and $C_6H_4$ prevented their detection. However, their presence could not be confirmed for the May 2015 period and both compounds were considered to be absent or below the estimated detection limit. Further investigations were made regarding the presence of glycerol at 92 u/e. Several investigations of glycerol in the comet 67P have been performed previously. Altwegg et al.[7] suggested the presence of toluene instead of glycerol. In fact, the mass/charge ratio of toluene and glycerol are comparable, even the measurements in the DFMS high-resolution mode were not sufficient for clear separation of both peaks. The calibration of glycerol preformed in this study; however, revealed that fragmentation behavior of glycerol is not consistent with the data from May 2015, as major fragments are missing. This confirms the findings of Altwegg et al.[7], suggesting the peak at 92 u/e is due to toluene instead. Table 2 provides a list with all CHO-bearing molecules detected in May 2015.

**Table 2: CHO-bearing molecules present in May 2015**

| Integer mass (u) | compound | Interpretation |
|---|---|---|
| 29 | COH | Fragment |
| 30 | CH2O | Formaldehyde |
| 31 | CH3O | Fragment |
| 32 | CH4O | Methanol |
| 33 | CH5O | Fragment |
| 41 | C2HO | Fragment |
| 42 | C2H2O | Fragment |
| 43 | C2H3O | Fragment |
| 44 | C2H4O | Acetaldehyde |
| 45 | C2H5O | Fragment |
| 46 | CH2O2 | Formic Acid |
| 46 | C2H6O | Ethanol |
| 57 | C3H5O | Fragment |
| 58 | C3H6O | Propanal or Acetone* |
| 60 | C2H4O2 | Acetic Acid |
| 62 | C2H6O2 | Ethylene Glycol |
| 70 | C4H6O | Fragment |
| 71 | C4H7O | Fragment |
| 72 | C4H8O | Butanal* |
| 74 | C3H6O2 | Methyl Acetate* |

\* Molecular identification is based on mass/charge ratios, hence the definition of the exact structural group is difficult as several stable or semi-stable modifications with the same set of elements might appear. Isomerism may occur but cannot be considered here. Therefore, the names of the species identified here may not be unique unless identified in earlier studies; see also [4,7].

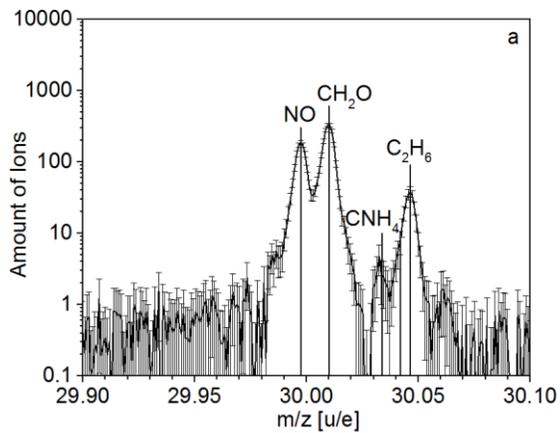
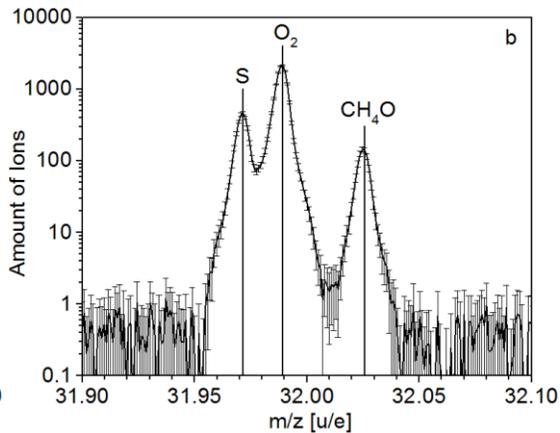
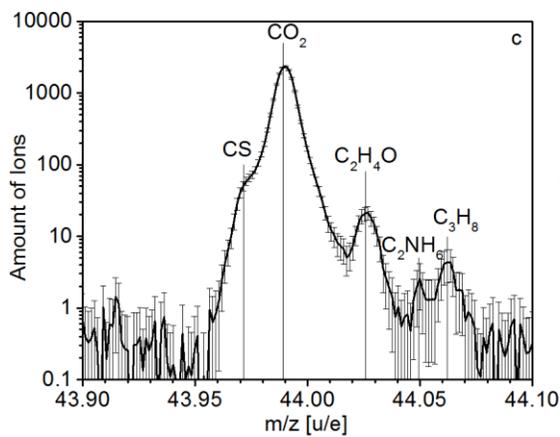
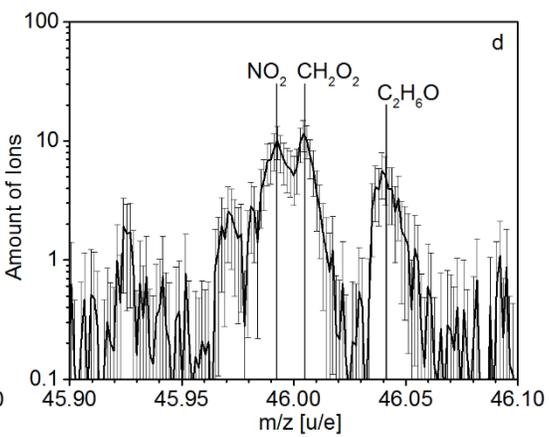
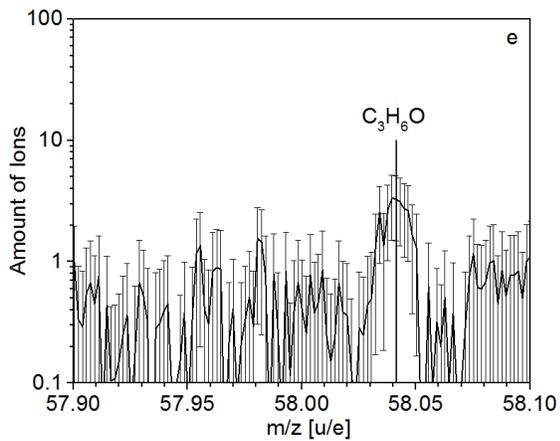
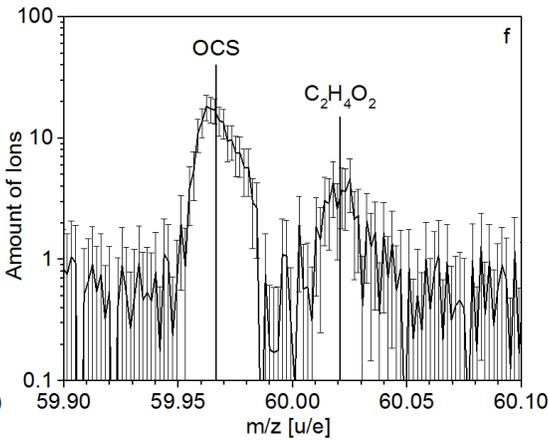

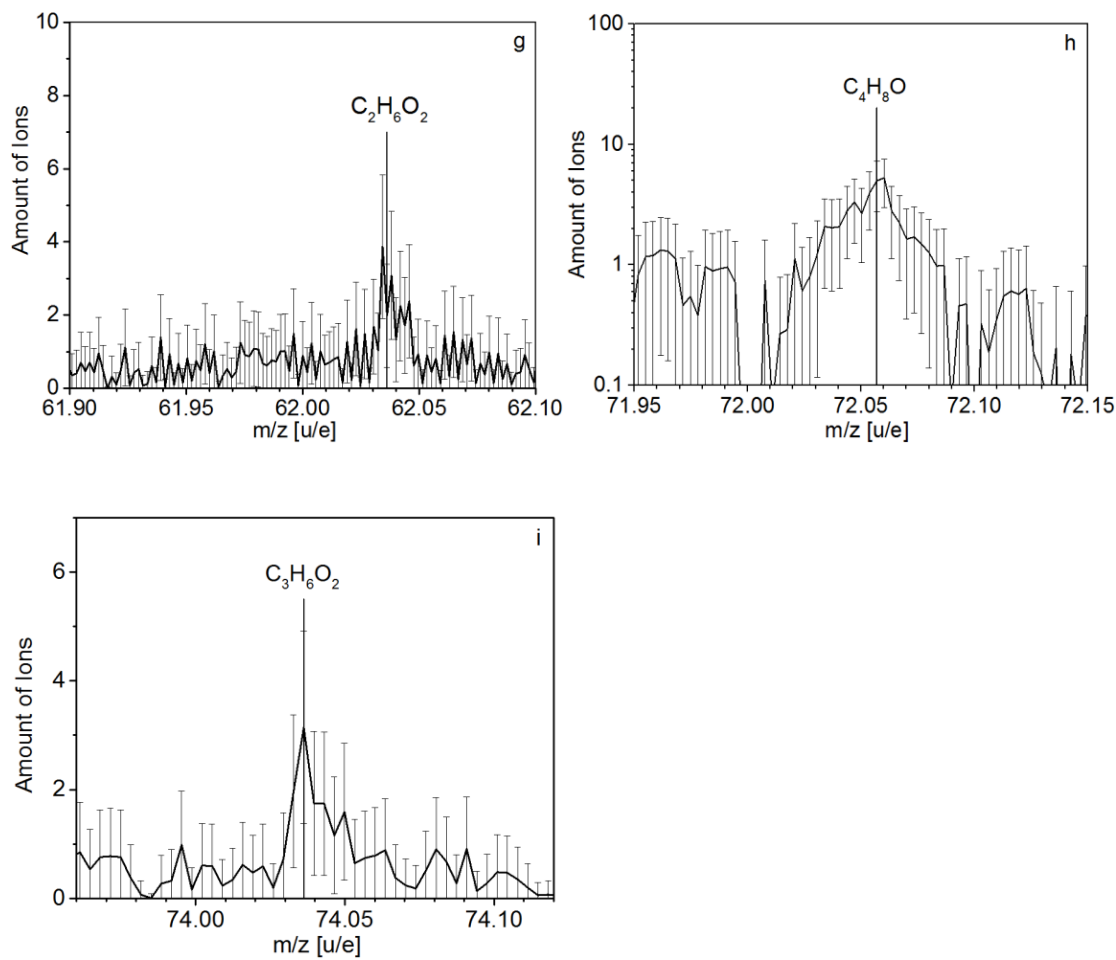

Figure 3.a-i: DFMS spectra of CHO-bearing molecules detected in May 2015 post-equinox-period up to mass 100 u. 3.a: Formaldehyde; 3.b: Methanol; 3.c: Acetaldehyde; 3.d: Formic Acid, Ethanol; 3.e: Propanal or Acetone; 3.f: Acetic Acid; 3.g: Ethylene Glycol; 3.h: Butanal; 3.i: Methyl Acetate;

3.2 Bulk Abundances of CHO-bearing Molecules

Studies of the abundances of CHO-bearing compounds using ground-observations can be found for most types of comets. Also, Le Roy et al.[4] calculated abundances relative to water for the comet 67P at larger heliocentric distance in 2014, shortly after the spacecraft's arrival. Table 3 lists the abundances relative to water for all CHO-molecules detected up to 100 u/e in the May 2015 post-equinox period. Formaldehyde and methanol show the highest abundances. Their abundances of 0.32% and 0.21%, respectively are within 2σ of each other. Other CHO-molecules appear to be lower by approximately a factor of 10 compared to methanol and formaldehyde. They cover a relatively small range from 0.01% to 0.05% relative to water. In this second group, ethanol and acetaldehyde show relatively high abundances of close to 0.05%. Methyl acetate, propanal (or acetone), acetic acid and butanal show much lower abundances of less than 0.01% relative to water.

**Table 3: Abundances of CHO-bearing molecules (rel. to $H_2O$)**

| Integer Mass (u) | CHO-molecule | Abundance |
|---|---|---|
| 30 | Formaldehyde | $(3.2 \pm 1.0) \cdot 10^{-3}$ |
| 32 | Methanol | $(2.1 \pm 0.6) \cdot 10^{-3}$ |
| 44 | Acetaldehyde | $(4.7 \pm 1.7) \cdot 10^{-4}$ |
| 46 | Formic acid | $(1.3 \pm 0.8) \cdot 10^{-4}$ |
| 46 | Ethanol | $(3.9 \pm 2.3) \cdot 10^{-4}$ |
| 58 | Propanal (Acetone) | $(4.7 \pm 2.4) \cdot 10^{-5}$ |
| 60 | Acetic acid | $(3.4 \pm 2.0) \cdot 10^{-5}$ |
| 62 | Ethylene glycol | $(1.1 \pm 0.7) \cdot 10^{-4}$ |
| 72 | Butanal | $(9.9 \pm 3.2) \cdot 10^{-5}$ |
| 74 | Methyl acetate | $(2.1 \pm 0.7) \cdot 10^{-5}$ |
| 18 | Water | 1.00 |

3.3 Comparison to Other Comets and the ISM

Abundances of methanol and formaldehyde have been summarized broadly for several types of comets. Most recently, Bockelée-Morvan and Biver[17] published a review of molecular abundances in cometary atmospheres, in which values of several CHO-bearing molecules are summarized from remote sensing spectroscopic observations. Formaldehyde and Methanol show the highest abundances, which is in agreement with our study, but broadly range between 0.6 - 6.2 % (methanol-to-water) and 0.13-2.4 % (formaldehyde-to-water). For other CHO-compounds, the upper limit appears to be significantly lower, e.g. 0.18 % for formic acid and 0.12% for ethanol. Bockelée-Morvan and Biver[17] further point out the differences of molecular abundances between various types of comets. This is in agreement with values we found in further existing literature: For comet Halley (1P) the ratios of methanol to water were found to be 1.8% by Bockelée-Morvan et al.[18] and Eberhardt et al.[19], and 1.7 % by Rubin et al.[20]. The ratio of formaldehyde to water ranges in Halley-type comets from 1.5% [20] to 4% [18]. Furthermore, methanol-to-water ratios for several long-period comets range from 1.48% in comet C2/2012 F6 [21], to 3.9% in comet C/2001 A2 [22].

For the majority of long-periodic comets methanol-to-water ratios around 2.5% were derived, such as for comet Hale-Bopp (C/1995) by Bockelée-Morvan et al.[23]. For these long period comets, formaldehyde-to-water ratios are generally lower, around 1% for comet Hale-Bopp [23] and for comet Hyakutake (C/1996-B2) [24].

The ratios of several other CHO-bearing molecules to water, including formic acid, acetaldehyde, and ethylene glycol, have been determined for long-periodic comets. Generally, the highest abundances for ethylene glycol-to-water are around 0.3%: Comet Hale-Bopp (0.25%) [25], comet

Lemmon (0.24%), and comet Lovejoy (0.35%) [26]. Besides these molecules, most other CHO-bearing molecules in long periodic comets show abundances lower than 0.1% with respect to water. Relevant for this study are ratios derived from Jupiter-family comets, given similar short periodic orbit and evolutionary conditions to the comet 67P. For those comets, relative abundance determinations for minor CHO-bearing molecules are rare. Le Roy et al.[4] performed a study on the comet 67P, revealing abundances of 0.01% or lower for minor CHO-bearing molecules.

Like the long period comets, the abundances of methanol and formaldehyde in Jupiter-family comets are discussed broadly in the literature. Dello Russo et al.[27] provides a detailed and recent overview on molecular abundances of CHO-bearing molecules relative to water in Jupiter-family comets. The ratios appear to vary significantly among different Jupiter-family comets. The average methanol-to-water ratios range from 0.49 % in comet 73P/Schwassmann-Wachmann-C and 0.54 % in 73P/Schwassmann-Wachmann-B up to 3.48 % in comet 2P/Encke. Values available for comet 67P in 2014 show ratios on the lower side of this range, 0.31 to 0.55%, depending on the comet hemisphere.[4] The abundance of methanol in May 2015 is with 0.21%, hence low compared to the 2014 measurement and low compared to other Jupiter family comets. The differences between our results and Le Roy et al.[4] implies changes in the ratios relative to water over the Rosetta investigation period. This is caused by changing conditions when the spacecraft first encountered the comet and the post-equinox and pre-perihelion period in May 2015 analyzed here. In particular, Le Roy et al.[4] noted differences among the hemispheres of the comet and they used data from a period when the comet was beyond 3.1 au from the Sun and when the southern hemisphere was in winter. Furthermore, among others Hässig et al.[28], Fougere et al.[29], Läuter et al.[30], and Biver et al.[31] showed how relative abundances of a multitude of species varied as a function of heliocentric distance and spacecraft location at 67P.

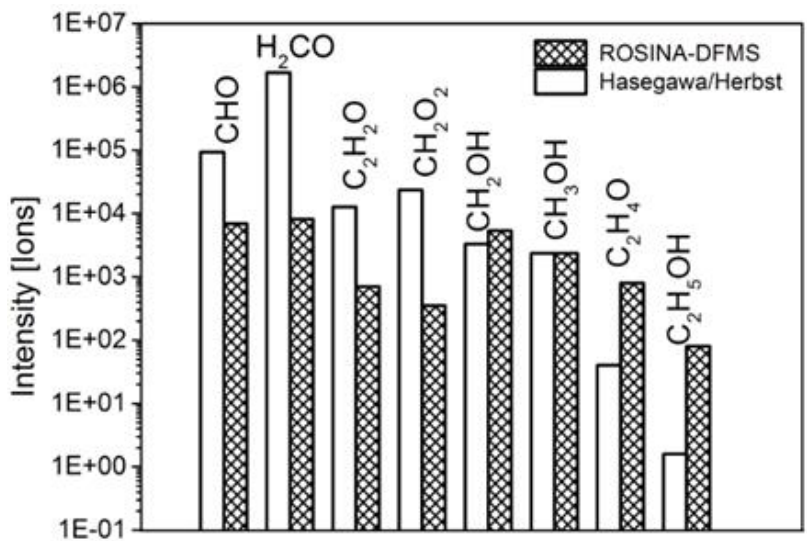

**Figure 4. Comparison of Hasegawa/Herbst ISM model and ROSINA-DFMS.** The plot shows the intensity of CHO-bearing molecules in the ISM modelled by Hasegawa and Herbst (1992) in comparison to the ROSINA-DFMS data from May 2015 at the comet 67P.

In earlier ROSINA studies such as Altwegg et al.[7], the presence of large amounts of unsaturated species was discussed. Furthermore, this study shows that fragmentation processes in the ion source are not sufficient for production of this amounts of molecules with unsaturated C=O bonds occurring in form of aldehydes and carboxylic acids. The presence of these compounds is important as it might provide hints at a common origin of the material in comets and the ISM. Hasegawa and Herbst[32] show models of gas-grain chemistry in dense interstellar clouds with complex organic molecules. Besides a high number of different organic species investigated in this study, some CHO-bearing molecules were taken into account. Figure 4 shows a comparison of the CHO-bearing molecules investigated in this study and in the Hasegawa/Herbst model (using results from Model A from Hasegawa and Herbst[32]). The values are normalized to methanol ($CH_3OH$) for better comparison and isomers occurring in the Hasegawa/Herbst model are appended to the compounds of same mass. Comparison of the Hasegawa/Herbst model to the ROSINA-DFMS results confirms that the investigated compounds are present in both studies, including saturated and unsaturated molecules. As can be seen in Figure 4, results from the Hasegawa/Herbst model show higher amounts for CHO, $H_2CO$, and $CH_2O_2$, with formaldehyde being especially enriched compared to ROSINA. $CH_2HO$ has similar abundance, while formic acid ($C_2H_4O$) and ethanol ($C_2H_5OH$) are more abundant in the ROSINA data.

## 4. Conclusions

A laboratory calibration campaign of ROSINA-DFMS on CHO-bearing molecules was conducted to determine the molecular fragmentation pattern and sensitivity of the instrument to the molecules. Based on these lab results, a detailed identification and quantification campaign of CHO-bearing molecules in the DFMS space data for May 2015 post-equinox period was performed. This campaign revealed the presence of CHO-bearing molecules most-likely in the form of aldehydes, alcohols, and carboxylic-acids. The abundances of all compounds present were calculated relative to water. Formaldehyde and methanol are the most abundant of all CHO-bearing molecules present, followed by acetaldehyde and ethanol. Other molecules present are significantly less abundant and have a maximum of 0.01 % abundance relative to water. The ratios of methanol and formaldehyde with respect to water are similar to values derived for other Jupiter-Family comets but are at the lower end of the range. The abundances are generally lower than the ones derived by Le Roy et al.[4] for comet 67P using DFMS data from early in the encounter when the comet was far from the Sun and relative water outgassing was lower.[cf. 33] The observations here are exclusively from a period when the comet was much closer to the Sun and the southern hemisphere was in summer.

The time period here is closer to perihelion, when most ground-based observations of comets occur. Therefore, based on the comparison with ground-based measurements of other comets, 67P is classified as relatively poor in CHO-bearing species. Therefore, it is important that there are future rendezvous with other Jupiter-family comets and detailed comparison between in situ and ground-based observations. Similarly, ground-based observations of comets far from the perihelion would help confirm the differences in the in situ measurements of CHO-bearing species.

Finally, concerning CHO-bearing species in comets and the ISM, abundant unsaturated CHO-bearing species were detected in form of aldehydes and carboxylic acids with unsaturated C=O bonds in comet 67P. A comparison to the model of Hasegawa and Herbst[32], showing the evolution of organics in the ISM, highlights the similarities to the ROSINA data from comet 67P on CHO-bearing molecules, with a difference in abundance especially for formaldehyde. This difference might indicate the conservation of CHO-compounds in comets from earlier solar system formation stages.

SUPPORTING INFORMATION

Appendix A: Table with Sensitivity and Fragmentation values as used for this work;

Appendix B: Fragmentation patterns as used for this work.

ACKNOWLEDGMENT

This work was supported by the following institutions and agencies: University of Bern was funded by the State of Bern, the Swiss National Science Foundation (SNSF, 200021_165869 and 200020_182418) and by the European Space Agency PRODEX Program. Work at Southwest Research institute was funded by Jet Propulsion Laboratory (subcontract no. 1496541), at the University of Michigan by NASA (contract JPL-1266313), by CNES grants at Laboratoire Atmosph`eères, Milieux, Observations Spatiales, and at the Royal Belgian Institute for Space Aeronomy by the Belgian Science Policy Office via PRODEX/ROSINA PEA 90020. Further we would like to thank all the engineers, technicians and scientists involved in the mission, the Rosetta Spacecraft, and the ROSINA team. Rosetta is an ESA mission with contributions from its member states and NASA. Thus we acknowledge herewith the work of the entire ESA Rosetta team. All ROSINA data have been released to the public PSA archive of ESA (https://www.cosmos.esa.int/web/psa/rosetta) and to the PDS archive of NASA.

AUTHOR INFORMATION

*Corresponding Author:

Markus Schuhmann

Sidlerstrasse 5, 3012 Bern, Switzerland

Email: Markus.Schuhmann@space.unibe.ch

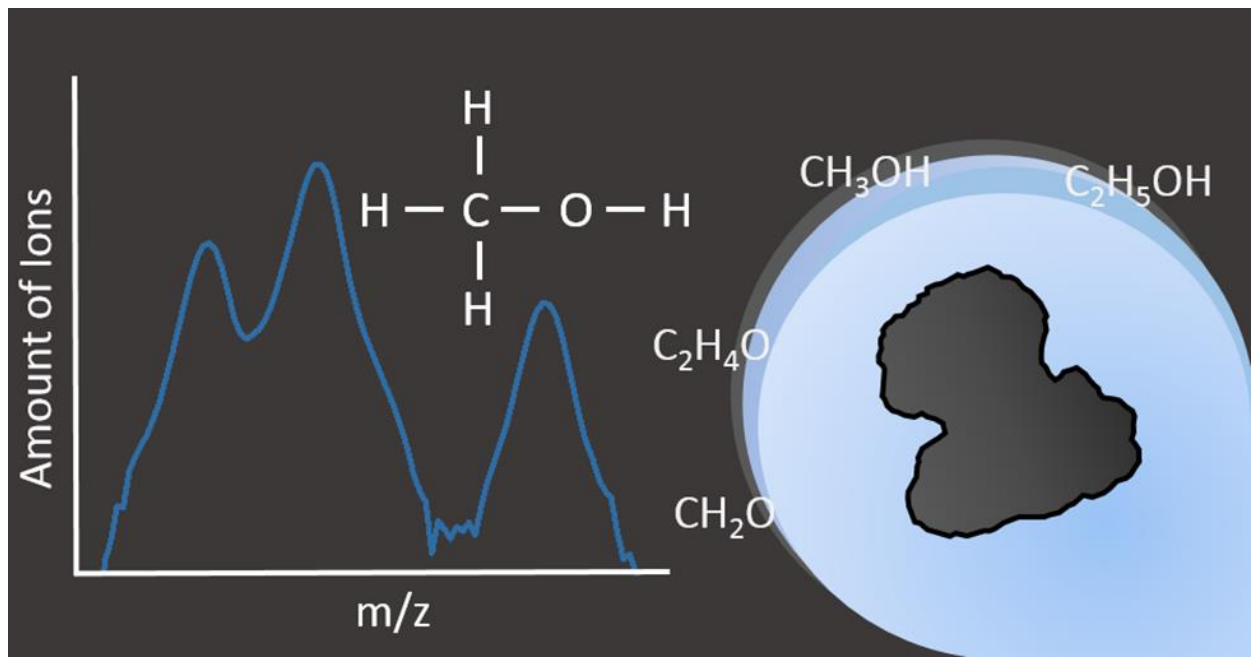